\documentclass[conference]{IEEEtran}
\IEEEoverridecommandlockouts
\usepackage{cite}
\usepackage{amsmath,amssymb,amsfonts}
\usepackage{algorithmic}
\usepackage{algorithm}
\usepackage{graphicx}
\usepackage{textcomp}
\usepackage{xcolor}
\usepackage[hidelinks]{hyperref}
\usepackage[a4paper, total={184mm,239mm}]{geometry}
\def\BibTeX{{\rm B\kern-.05em{\sc i\kern-.025em b}\kern-.08em
    T\kern-.1667em\lower.7ex\hbox{E}\kern-.125emX}}
\begin{document}

\title{AERO: Adaptive and Efficient Runtime-Aware OTA Updates for Energy-Harvesting IoT}

\author{
    \IEEEauthorblockN{
        Wei Wei\IEEEauthorrefmark{1}, 
        Jingye Xu\IEEEauthorrefmark{1}, 
        Sahidul Islam\IEEEauthorrefmark{2}, 
        Dakai Zhu\IEEEauthorrefmark{1}, 
        Chen Pan\IEEEauthorrefmark{3}, 
        Mimi Xie\IEEEauthorrefmark{1}
    }
    \IEEEauthorblockA{\IEEEauthorrefmark{1}Department of Computer Science, The University of Texas at San Antonio, San Antonio, TX, USA}
    \IEEEauthorblockA{\IEEEauthorrefmark{2}Department of Computer Science, Kennesaw State University, Marietta, GA, USA}
    \IEEEauthorblockA{\IEEEauthorrefmark{3}Department of Electrical and Computer Engineering, The University of Texas at San Antonio, San Antonio, TX, USA}
    \small E-mail: \{wei.wei2, jingye.xu, dakai.zhu, chen.pan, mimi.xie\}@utsa.edu, sislam19@kennesaw.edu
}

\maketitle

\begin{abstract}
Energy-harvesting (EH) Internet of Things (IoT) devices operate under intermittent energy availability, which disrupts task execution and makes energy-intensive over-the-air (OTA) updates particularly challenging. Conventional OTA update mechanisms rely on reboots and incur significant overhead, rendering them unsuitable for intermittently powered systems. Recent live OTA update techniques reduce reboot overhead but still lack mechanisms to ensure consistency when updates interact with runtime execution. This paper presents AERO, an Adaptive and Efficient Runtime-Aware OTA update mechanism that integrates update tasks into the device’s Directed Acyclic Graph (DAG) and schedules them alongside routine tasks under energy and timing constraints. By identifying update-affected execution regions and dynamically adjusting dependencies, AERO ensures consistent update integration while adapting to intermittent energy availability. Experiments on representative workloads demonstrate improved update reliability and efficiency compared to existing live update approaches.
\end{abstract}


\section{Introduction}
\label{introduction}

The growing demand for sustainable, low-maintenance computing has accelerated the adoption of EH technologies in next-generation IoT devices. By harvesting energy from ambient sources such as solar, thermal, kinetic, and radio-frequency signals \cite{sun2023requirements}, EH devices eliminate the need for frequent battery replacement and enable autonomous operation in resource-constrained environments. These capabilities make EH systems particularly valuable in applications such as agricultural sensing \cite{aggarwal2024studies}, medical implants \cite{shuvo2022energy}, wildlife tracking \cite{toledo2022vildehaye}, and disaster recovery \cite{sun2022sequential}. The global market for EH technologies is projected to reach \$6.5 billion by 2028 \cite{market_size}, underscoring their growing role in future sustainable IoT infrastructure.

To ensure the long-term functionality, security, and adaptability of EH devices, OTA updates provide a promising mechanism for firmware maintenance in EH IoT systems. However, frequent power interruptions create unpredictable execution windows, making it difficult to safely and efficiently complete OTA updates. Traditional OTA update mechanisms \cite{dong2012r2, wei2022intermittent} typically rely on rebooting into a bootloader to apply firmware updates. In this process, update packets are received and stored in non-volatile memory during normal execution, after which the system reboots to install the update. On devices with limited memory, updates must often be processed in smaller chunks, leading to multiple reboots. Each reboot incurs substantial energy overhead, increases latency, and disrupts routine execution. For intermittently powered EH devices, these cumulative costs significantly prolong downtime and increase the risk of incomplete or failed updates.

Live OTA update techniques enable updates to be applied at runtime without rebooting into a bootloader, thereby improving execution continuity \cite{zhang2016live}. While this approach reduces downtime and preserves system availability, it introduces new correctness challenges. During live updates, updated and non-updated code regions may temporarily coexist, and residual state in volatile memory can become inconsistent with the new logic. Such cross-version execution can compromise correctness and stability, motivating the need for mechanisms that ensure updates are applied only at safe execution points.

Although more recent live OTA update techniques have been proposed for EH IoT devices \cite{wei2024intermittent}, they still fall short in ensuring correctness during execution. Existing approaches either overlook consistency issues that arise at runtime or avoid them by postponing updates until all routine tasks have completed. This limitation motivates the need for a runtime-aware OTA update mechanism that explicitly accounts for execution state.

% To address these limitations, we propose AERO, which coordinates update execution with ongoing tasks to preserve correctness under intermittent energy. Upon receiving an update request, AERO adjusts the DAG by integrating update tasks and their dependencies with existing tasks. Based on the resulting DAG, AERO then coordinates the execution of update and routine tasks under energy and timing constraints at runtime.

To address these limitations, we propose AERO, a runtime-aware OTA update mechanism for EH IoT devices. The key contributions are summarized as follows:

\begin{itemize}
% \vspace{-12pt}
\item We propose a lightweight OTA packet format that encodes update task relations, routine task dependencies, and update operations for both update code and the DAG.  

\item We formally define \textit{mutually dependent update groups} and use them to identify the corresponding \textit{update-affected block} in the DAG, which captures update consistency requirements.

\item We devise a runtime DAG adjustment algorithm that integrates update tasks and their dependencies with routine tasks under different update scenarios, enabling coordinated execution during updates.

\item We propose a unified scheduling method that jointly schedules update and routine tasks under intermittent energy availability and deadline constraints.

\end{itemize}
\section{Background and Motivation}
\label{background}

\subsection{Background}

Several studies \cite{koshy2005remote, han2005sos, panta2009hermes, dong2012r2, wei2022intermittent, wei2024intermittent} have explored OTA update techniques for resource-constrained IoT devices, many of which are applicable to EH IoT devices. Much of this work reduces transmission and energy overhead through smaller update sizes, improved code similarity, and optimized encoding. Other efforts employ checkpointing strategies \cite{singla2022survey, Shyamala, Sahidul} to tolerate power interruptions, or hardware-aware methods \cite{wei2024energy, banerjee2021memory, liu2023light, sun2024flora+, jewsakul2024fiora} that align updates with available energy and memory constraints.

To provide context for our approach, we review three representative categories of OTA update techniques: whole image, incremental, and live update.

\subsubsection{Whole Image Update}

Whole image update replaces the entire firmware image by transmitting it to the device and overwriting the existing one. MSP430FRBoot \cite{Brown_Pier_Gao_2020} implements this approach using dual-flash banks, where the new image is written to an inactive bank and activated upon completion. Light Flash Write \cite{liu2023light} improves reliability under intermittent power by lowering the energy cost of flash writes. However, whole image update remains costly for EH IoT devices, as transferring and writing an entire firmware image requires significant energy and memory resources, increasing the likelihood of incomplete updates under intermittent execution.

\subsubsection{Incremental Update}

Incremental update reduces overhead by transmitting only modified firmware portions using delta encoding \cite{wei2022intermittent}. For flash-based devices, segment-based designs minimize memory operations and energy usage during intermittent updates \cite{wei2024energy}. At the network level, multicast and beamforming improve transmission efficiency across multiple devices in LoRa networks \cite{jewsakul2024fiora, sun2024flora+}. Despite these benefits, incremental update still relies on rebooting into a bootloader, which is problematic for intermittently powered devices.

\subsubsection{Live Update}

Live update applies firmware modifications during runtime without requiring reboots. Trampoline-based patching \cite{zhang2016live} introduced atomic code replacement, and more recent work extends this capability to intermittently powered devices \cite{wei2024intermittent}. While these techniques reduce reboot overhead, they often assume that updates can be safely applied without causing inconsistencies, which is not always valid in real-time or intermittently powered systems.

\subsection{Motivation}

\begin{figure}[h]
    \centering
    \includegraphics[width=8.2cm]{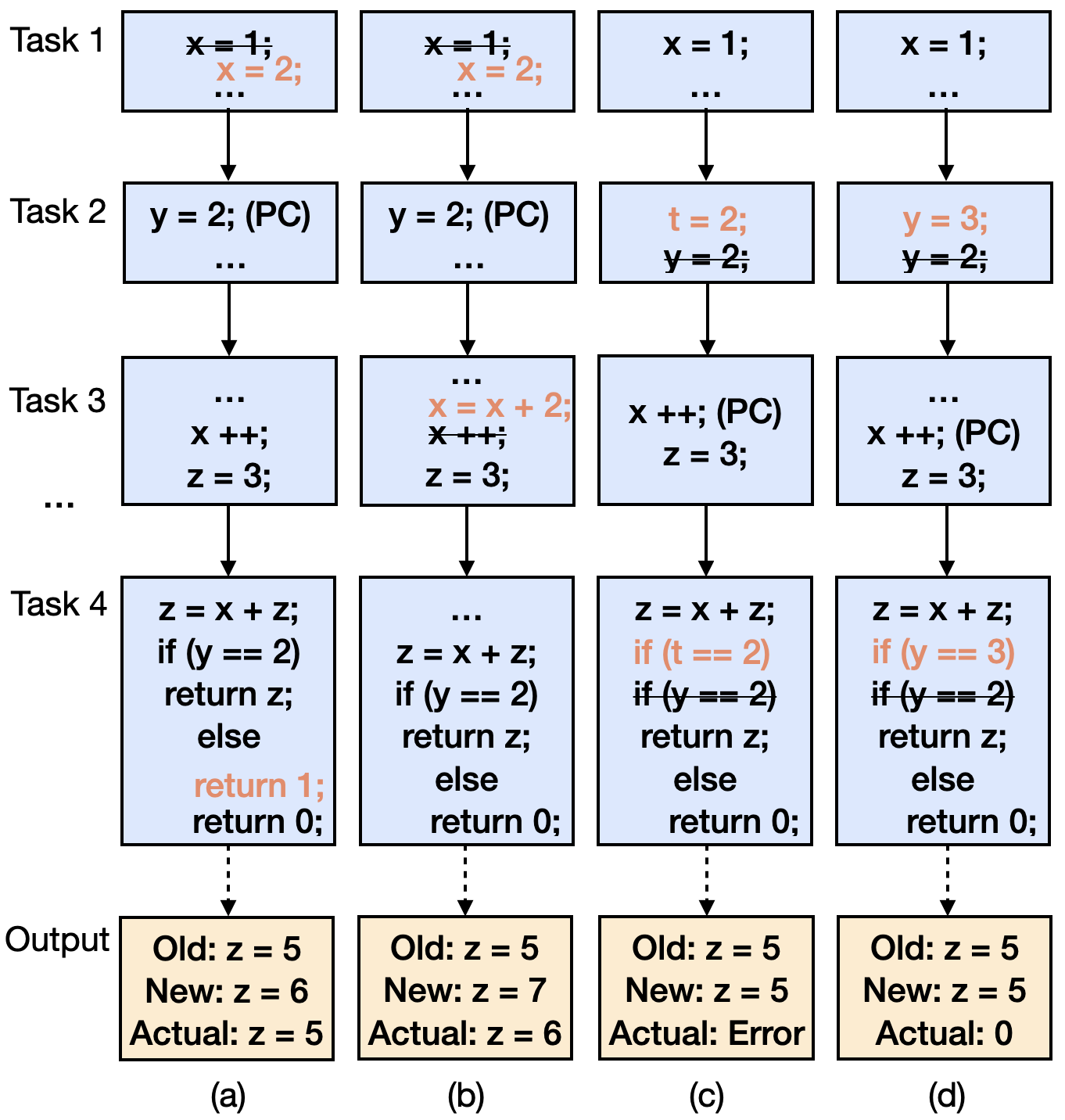}
    \vspace{-6pt}
    \caption{Motivating Examples of Mixed-Version Execution: (a) No Impact, (b) Incorrect Computation, (c) Execution Failure, (d) Unintended Interaction.}
    \label{fig_motivation}
\end{figure}

Although live OTA updates reduce reboot overhead, they also introduce new consistency risks, highlighting the need for runtime-aware OTA update mechanisms that explicitly manage interactions between update tasks and ongoing execution. During runtime modification, updated and non-updated code regions may temporarily coexist, leading to mixed-version execution. In some cases this has no observable effect, but in others it can cause incorrect computation, execution failure, or unintended interactions, as illustrated in Fig.~\ref{fig_motivation}. These risks are amplified in EH devices, where firmware is decomposed into fine-grained tasks to align with short power cycles, sometimes only a few instructions long \cite{xu2024cache, li2016performance}. Under such conditions, even subtle inconsistencies can propagate into diverse failure types. Although prior work \cite{wei2024intermittent} acknowledges these issues, the challenge of handling mixed-version execution remains unresolved.

A further motivation comes from the observation that update tasks do not need to strictly wait until routine tasks complete, nor should routine tasks be stalled until updates finish. Task execution in EH devices often contains idle periods due to scheduling gaps, during which update tasks can safely proceed. By treating updates as schedulable entities within the DAG, such idle time can be effectively utilized to advance updates without disrupting routine execution.

In addition to scheduling concerns, updates may also require structural changes to the DAG itself. Some updates modify tasks or dependencies, add new tasks, or remove obsolete ones, thereby reshaping the DAG beyond simple code replacement. To the best of our knowledge, no existing OTA update approach for EH IoT devices supports such DAG-level updates.
\section{AERO: Runtime-Aware OTA Update Mechanism}
\label{technique}

We develop AERO, which integrates update tasks and their dependencies into the runtime DAG using a lightweight, dependency-driven packet format. AERO performs runtime DAG adjustment followed by unified scheduling to coordinate update and routine tasks under energy and timing constraints. AERO enforces task-level control-flow and dependency consistency, assuming tasks encapsulate local state and access shared memory, peripherals, and interrupts only at task boundaries.

% last sentence try to exclude those complicated data dependency cases.

\subsection{Definition}
\label{tech_1}

\subsubsection{Mutually Dependent Update Group Definition}

Let \(T = T_{\text{old}} \cup T_{\text{new}}\) denote all firmware tasks, where 
\(T_{\text{old}} = \{ t_1, t_2, \dots, t_m \}\) are existing tasks and \(T_{\text{new}} = \{ t_{m+1}, t_{m+2}, \dots, t_n \}\) are new tasks introduced by updates. Execution is modeled as a DAG \( G = (T, E) \), where \(E \subseteq T \times T\) and \((t_i, t_j) \in E\) indicates that \(t_j\) depends on \(t_i\). Correspondingly, let \( U = U_{\text{old}} \cup U_{\text{new}} \) denote the set of update tasks, with each \( u_i \in U \) associated with its corresponding \( t_i \in T \). Update tasks modify both the code of their associated tasks and any linked dependencies. An update task \( u_i \) depends on another update task \( u_j \) if both must be updated in the same process for correctness. A \textit{mutually dependent update group} is a minimal subset \(M \subseteq U\) where every \(u_i \in M\) depends on at least one \(u_j \in M\), and correctness is preserved only when all tasks in \(M\) are updated together.

\subsubsection{Update-Affected Block Definition}

The \textit{update-affected block} is the minimal subgraph of the DAG containing all tasks in a \textit{mutually dependent update group}, along with intermediate nodes and edges on the dependency paths connecting them. This block represents the execution region influenced by the update and must be treated as a whole during integration.

\subsection{Dependency-Driven Update Packet}
\label{tech_2}

\begin{figure}[h]
    \centering
    \includegraphics[width=7cm]{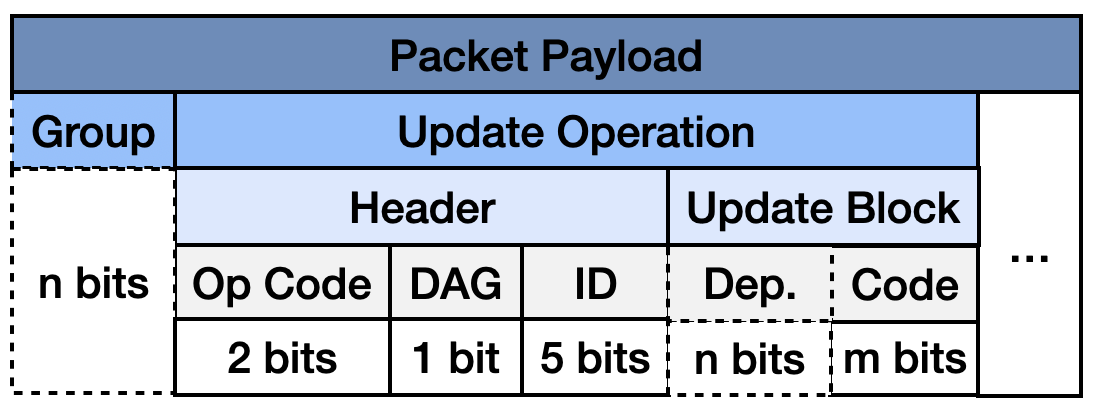}
    \vspace{-6pt}
    \caption{AERO Update Packet Structure}
    \label{fig_packet}
\end{figure}

AERO introduces a dependency-driven update packet format, shown in Fig.~\ref{fig_packet}, to enable runtime-aware integration of updates in EH IoT devices. Each packet payload consists of a group field and an update operation. The group field, included only in the first packet, encodes the \textit{mutually dependent update group} with one bit per routine task in the DAG, where 1 indicates association with the update and 0 indicates none. Unused positions are reserved for future tasks. This field allows the system to determine the \textit{update-affected block} early and supports incremental processing, so that updates can progress even when memory cannot hold the entire update at once.

The update operation contains a header and an update block. The header begins with a 2-bit operation code specifying the update type, based on intermittent-aware update operations in \cite{wei2024intermittent}. A 1-bit DAG flag indicates whether the update block carries dependency information. If set, the first $n$ bits of the update block form an optional dependency field, with each bit specifying whether the update task depends on the corresponding routine task. The remaining header bits identify the task ID, and the update block concludes with an $m$-bit code segment delivering the update content.

\subsection{Runtime DAG Adjustment}

\begin{figure}[h]
    \centering
    \includegraphics[width=7.5cm]{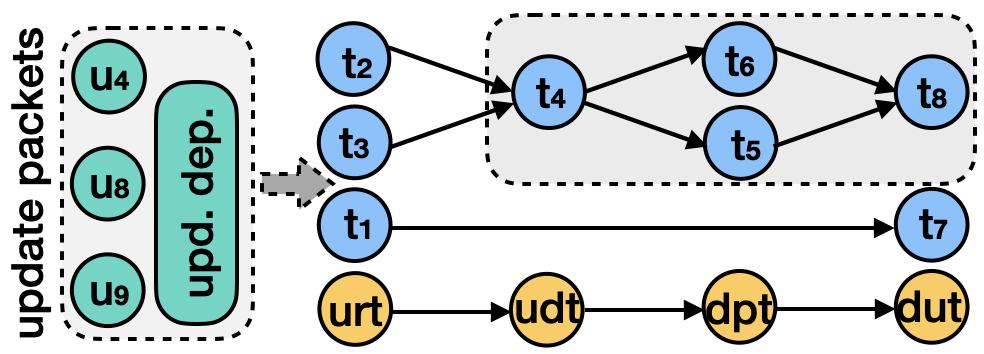}
    \vspace{-6pt}
    \caption{Overview of the AERO Update Process.\\
    \small Update Receiving Task (urt): collects incoming packets; Update Decoding Task (udt): decodes packets and reconstructs update tasks; Dependency Processing Task (dpt): identifies the \textit{update-affected block}; DAG Updating Task (dut): inserts update tasks and adjusts affected dependencies in $G$.}
    \label{fig_t0}
\end{figure}

To support runtime-aware updates without interrupting execution, AERO introduces a dedicated update task chain that is triggered upon receiving an update notification. As shown in Fig.~\ref{fig_t0}, the chain inserts the update tasks together with their dependency adjustments into the DAG.

\subsubsection{Modifying Existing Tasks}

\begin{figure*}[t]
    \centering
    \includegraphics[width=0.9\textwidth]{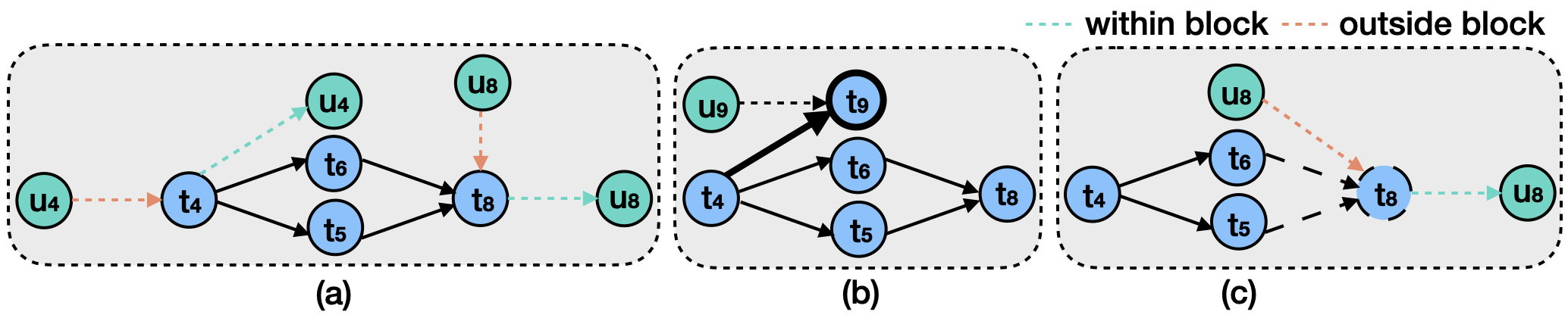}
    \vspace{-6pt}
    \caption{AERO Update Cases: (a) Modifying Existing Tasks $t_4$ and $t_8$; (b) Inserting New Task $t_9$; (c) Removing Obsolete Task $t_8$}
    \label{fig_t1}
\end{figure*}

The framework supports updating both tasks and their dependencies when modifying existing tasks. In the example of Fig.~\ref{fig_t1}(a), update tasks modify $t_4$ and $t_8$. Integration depends on the current execution. If runtime is within the \textit{update-affected block} (e.g., at $t_4$, $t_5$, $t_6$, or $t_8$), updates are deferred until the corresponding tasks complete so the current cycle finishes under the old version. Otherwise, updates are inserted before their associated tasks and the system adopts the new version in the same execution. For clarity, this example shows only task updates without dependency changes.

\subsubsection{Inserting New Tasks} 

The framework also supports inserting new tasks into the DAG. In Fig.~\ref{fig_t1}(b), update task $u_9$ creates a new task $t_9$. Because $t_9$ does not exist beforehand, $u_9$ must execute first regardless of runtime to ensure the task is available for scheduling. In practice, new tasks may introduce dependencies with existing tasks, requiring joint updates to preserve consistency. For clarity, this example shows only insertion of $t_9$ without dependency changes.

\subsubsection{Removing Obsolete Tasks}

The framework also supports removing obsolete tasks from the DAG. In Fig.~\ref{fig_t1}(c), update task $u_8$ removes task $t_8$ and all linked dependencies. If runtime is within the \textit{update-affected block}, removal is deferred until $t_8$ completes; otherwise it occurs before $t_8$ begins. In practice, a task may have dependencies with others, and its removal requires updating those tasks to maintain consistency. For clarity, this example shows only removal of $t_8$ without additional dependency changes.

\subsubsection{Runtime DAG Adjustment Algorithm}

Based on the illustrated cases, we now present AERO's algorithm for runtime DAG adjustment in EH IoT devices. The procedure first inserts a virtual start node $s$ connected to all original sources, ensuring that even if the \textit{update-affected block} begins at the graph entry, execution can still be temporarily blocked before entering the block. The block $B$ is then identified, and all incoming edges into $B$ are removed to prevent new execution paths while updates are pending. This blocking is essential when memory cannot hold all update packets, since without it routine tasks inside $B$ could execute under outdated code, leading to inconsistencies. By deferring entry into $B$ until update tasks are completed, AERO preserves correctness while allowing unaffected tasks outside the block to continue. Following prior intermittent update designs \cite{wei2024intermittent}, AERO checkpoints both update and routine tasks in non-volatile memory, allowing partially applied updates to be safely recovered and re-applied after power loss.

\begin{algorithm}[htbp]
\caption{Runtime DAG Adjustment Algorithm}
\label{alg_1}
\small
\begin{algorithmic}[1]
\STATE \textbf{Input:} $G=(T_{\text{old}},E)$; update group $M$; current task $t_{\text{curr}}$
\STATE \textbf{Output:} Modified DAG $G'$, blocked edges $E_{\text{blk}}$
\vspace{3pt}

\STATE $T' \gets T_{\text{old}} \cup \{s\}$,\quad $E' \gets E \cup \{(s, v) \mid v \in \mathrm{sources}(T_{\text{old}})\}$ 

\hfill // add virtual start node $s$

\STATE $B \gets f(M,(T',E'))$ \hfill // identify update-affected block
\STATE $E_{\text{blk}} \gets \{ (x,y) \in E' \mid y \in B \}$ \hfill // backup blocked edges
\STATE $E' \gets E' \setminus E_{\text{blk}}$ \hfill // block entry into $B$

\FOR{each $u_i \in M$}
    \STATE let $t_i$ be the task associated with $u_i$
    \IF{$t_i \notin T'$}
        \STATE $T' \gets T' \cup \{u_i, t_i\}$ \hfill // add $u_i$ and placeholder $t_i$
        \STATE $E' \gets E' \cup \{ (u_i, t_i) \}$ \hfill // add edge $(u_i,t_i)$
    \ELSE
        \IF{$t_{\text{curr}} \in B$}
            \STATE $T' \gets T' \cup \{u_i\}$ 
            \STATE $E' \gets E' \cup \{ (t_i, u_i) \}$ \hfill // defer until $t_i$ completes
        \ELSE
            \STATE $T' \gets T' \cup \{u_i\}$
            \STATE $E' \gets E' \cup \{ (u_i, t_i) \}$ \hfill // update before $t_i$
        \ENDIF
    \ENDIF
\ENDFOR

\STATE $G' \gets (T', E')$
\STATE \textbf{return } $G', E_{\text{blk}}$
\end{algorithmic}
\end{algorithm}

\subsection{Unified Scheduling Algorithm}
\label{tech_3}

\begin{figure}[h]
    \centering
    \includegraphics[width=8.5cm]{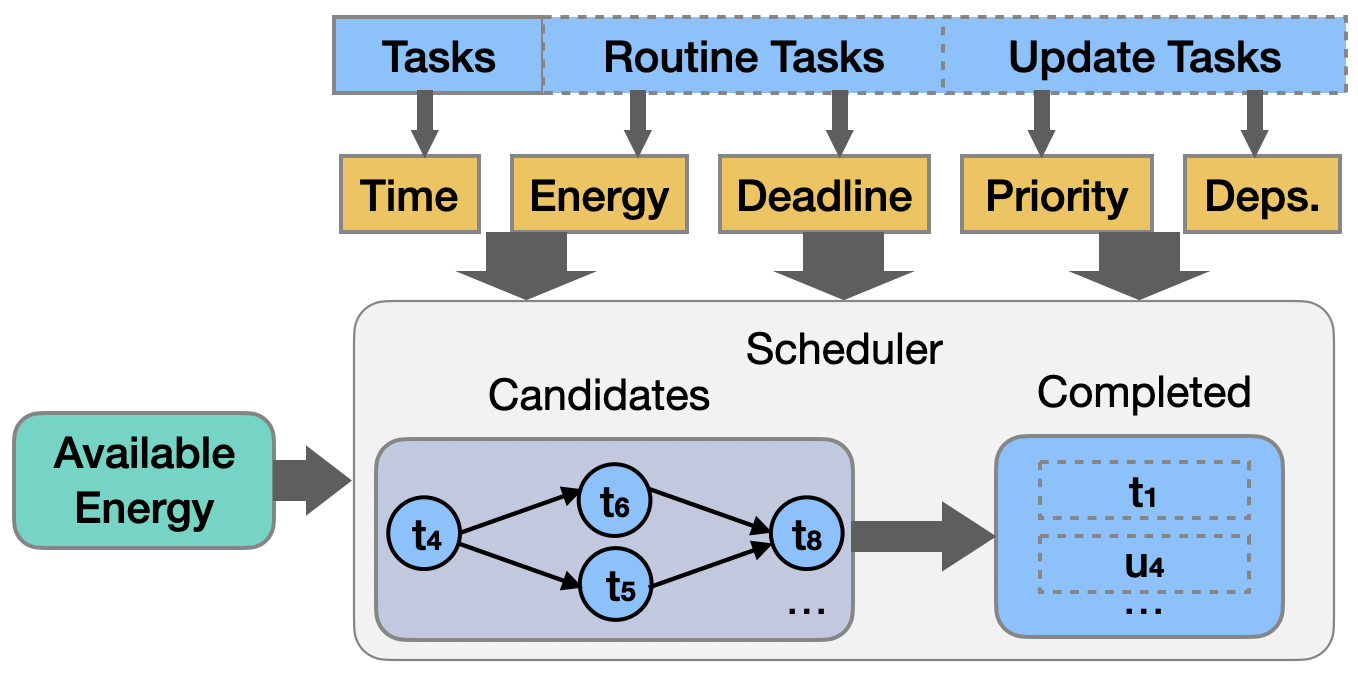}
    \caption{Graphical Illustration of Tasks Selection Process}
    \label{fig_scheduler}
\end{figure}

Once update tasks are integrated, AERO employs a unified scheduler to coordinate routine and update tasks while adapting to energy and deadline constraints. Figure~\ref{fig_scheduler} provides a graphical overview of the task selection process, while Algorithm~\ref{alg_2} presents the detailed procedure. Together, they show how updates are applied safely and efficiently without compromising real-time performance or exceeding energy budgets.

Task selection is handled with a priority queue ordered by deadline, then priority. Routine tasks execute directly, while update tasks are applied according to their operations. If the DAG bit in an update operation is set, the corresponding dependency edges are added to the DAG before the update is applied. When all update tasks in the current group $M$ are completed, the scheduler restores the entry edges of the \textit{update-affected block} $B$ removed in Algorithm~\ref{alg_1}. It then cleans the graph by removing update tasks, helper tasks, and the virtual start node along with their connected edges. The complete procedure is summarized in Algorithm~\ref{alg_2}.

\begin{algorithm}[htbp]
\caption{Unified Scheduling Algorithm}
\label{alg_2}
\small
\begin{algorithmic}[1]
\STATE \textbf{Input:} $G'=(T',E')$; blocked edges $E_{\text{blk}}$
\STATE \textbf{Output:} Executed task $t^*$ in each scheduling step
\vspace{3pt}

\STATE $Q \gets \emptyset$ \hfill // priority queue by deadline→priority

\WHILE{true}
    \FOR{each $t \in T'$ with $(\forall p \in \mathrm{pred}(t): p \text{ completed})$}
        \STATE $Q \gets Q \cup \{t\}$
    \ENDFOR

    \IF{$Q = \emptyset$}
        \STATE \textbf{continue} \hfill // no schedulable task, recheck
    \ELSE
        \STATE $t^* \gets \text{pop}(Q)$ \hfill // earliest deadline, then priority
        \IF{$t^* \in T$}
            \STATE $\text{exec}(t^*)$ \hfill // execute routine task
        \ELSE
            \IF{$\text{header}(t^*).\text{DAG} = 1$}
                \STATE $E' \gets E' \cup \Delta E(t^*)$ \hfill // add new edges
            \ENDIF
            \STATE $\text{apply}(t^*)$ \hfill // apply update operation
        \ENDIF
        \STATE $\text{status}(t^*) \gets \text{completed}$
    \ENDIF

    \IF{$M \subseteq \{\text{completed tasks}\}$} 
        \STATE $E' \gets E' \cup E_{\text{blk}}$ \hfill // restore blocked edges into $B$
        \STATE $S \gets M \cup \{urt,udt,dpt,dut,s\}$
        \STATE $T' \gets T' \setminus S$ \hfill // remove update, helpers, virtual node
        \STATE $E' \gets E' \setminus \{\, (u,v) \in E' \mid (u \in S) \lor (v \in S) \,\}$ 
        \STATE $G \gets (T',E')$ \hfill // store updated DAG
        \STATE \textbf{break}
    \ENDIF
\ENDWHILE
\end{algorithmic}
\end{algorithm}

\section{Experiment}
\label{experiment}

\subsection{Experimental Setup}

All experiments are conducted on the TI MSP430FR5994 microcontroller~\cite{msp430fr5994}, a resource-constrained device featuring ultra-low power operation and embedded FRAM, making it suitable for EH IoT systems. Execution time and energy consumption are obtained using TI EnergyTrace~\cite{ti_energytrace} in Code Composer Studio (CCS)~\cite{ccstudio} for standalone profiling of benchmark tasks. Update tasks are assigned execution time and energy values based on their respective update sizes. These profiled results are later used to evaluate task execution and update performance under AERO’s unified scheduling.

\subsection{Benchmark Design}

To evaluate the effectiveness of AERO, we use four benchmarks that represent diverse application types and DAG structures. For each benchmark, the capacitor size is set so that the stored energy is just sufficient to execute the task with the highest energy demand. This configuration drives operation near the system’s energy limits, providing realistic conditions for assessing runtime-aware updates. Table~\ref{table_benchmark} summarizes the benchmarks and their configurations.

The four benchmarks capture different aspects of runtime-aware updates relevant to AERO. 
Quick Sort (B1)~\cite{guthaus2001mibench} illustrates fine-grained scheduling with very small tasks. AES Encryption (B2)~\cite{Shyamala} exercises update behavior across parallel but functionally distinct hardware and software execution paths. LeNet-5 (B3)~\cite{islam2022eve} provides a strictly linear task pipeline, suitable for analyzing update propagation under ordered execution. Heart Rate Monitor (B4)~\cite{xu2024real} combines parallelism and recombination in a fork-join DAG, enabling evaluation of consistency under partially parallel execution.

\begin{table*}[t]
\centering
\caption{Evaluation Benchmarks}
\label{table_benchmark}
\begin{tabular}{|| l | c | c | c | p{7cm} ||}
\hline
\textbf{Benchmark} & \textbf{Size (B)} & \textbf{DAG Type} & \textbf{Cap. (mF)} & \textbf{Description} \\ \hline\hline
B1: Quick Sort~\cite{guthaus2001mibench} & 1,696 & Linear & 0.02 & Four sequential quick sort operations \\ \hline
B2: AES Encryption~\cite{Shyamala} & 3,975 & Parallel & 0.2 & AES encryption/decryption using HW/SW paths \\ \hline
B3: LeNet-5~\cite{islam2022eve} & 50,632 & Linear & 10 & Inference using an optimized LeNet-5 model \\ \hline
B4: Heart Rate Monitor~\cite{xu2024real} & 46,050 & Fork-Join & 1 & PPG-based heart rate estimation \\ \hline
\end{tabular}
\end{table*}

\subsection{OTA Update Scenarios}

\begin{table}[htbp]
\centering
\caption{OTA Update Scenarios}
\label{table_cases}
\begin{tabular}{|c|p{4.6cm}|c|}
\hline
\textbf{ID} & \textbf{Description} & \textbf{Size (B)} \\ \hline
1 & B1: Modify 2nd sort task & 280 \\ \hline
2 & B2: AES key to 192-bit (acc/DMA) & 130 \\ \hline
3 & B2: AES block to 256-bit & 162 \\ \hline
4 & B3: Final layer to FC & 292 \\ \hline
5 & B4: Add sensor support & 702 \\ \hline
6 & B4: Update HR model \& UART & 6246 \\ \hline
\end{tabular}
\end{table}

Together, these benchmarks provide the basis for evaluating runtime-aware updates across diverse application types and DAG structures. Building on this foundation, we design six update scenarios derived from the benchmarks in Table~\ref{table_benchmark}. The scenarios cover two categories of updates. Functional modifications change task behavior, such as adjusting sorting routines, replacing the final stage of a neural network, or introducing new signal-processing steps. Parametric modifications change configuration parameters without affecting task behavior, for example increasing AES key or block sizes, updating model weights, or changing UART settings.  

Each scenario introduces one or more update tasks that form a \textit{mutually dependent update group}, which must be integrated at runtime without violating DAG dependencies or compromising consistency. Update sizes span from small task-level changes to large firmware updates in EH IoT devices. Table~\ref{table_cases} summarizes the six scenarios and their update sizes.

We do not define separate scenarios for full image update or DAG update. Full image update can be performed by any approach and does not differentiate them, while DAG update is uniquely supported by AERO and is evaluated through feature comparison in Section~\ref{evaluation}.

\section{Evaluation}
\label{evaluation}

Evaluation is conducted through simulation using execution time and energy values obtained from standalone task profiling. Available energy is updated continuously using real-world solar traces from an EH IoT device~\cite{banerjee2024autotile}, and update arrivals follow a pseudo-random process to emulate stochastic behavior. Task priorities follow the DAG, where earlier tasks have higher priority, later tasks have lower priority, and update tasks always run last. We assume sufficient non-volatile storage and that all update packets arrive before integration. Although AERO can handle multi-packet, incremental updates, this assumption removes communication cost as a variable and enables a fair comparison focused on runtime execution. For comparison, we evaluate AERO against two representative baselines: the live update baseline corresponds to the trampoline-based live update approach in~\cite{zhang2016live}, while the intermittent update baseline corresponds to the checkpoint-assisted OTA framework in~\cite{wei2024intermittent}.

\subsection{Feature Comparison}

\begin{figure}[h]
    \centering
    \includegraphics[width=8cm]{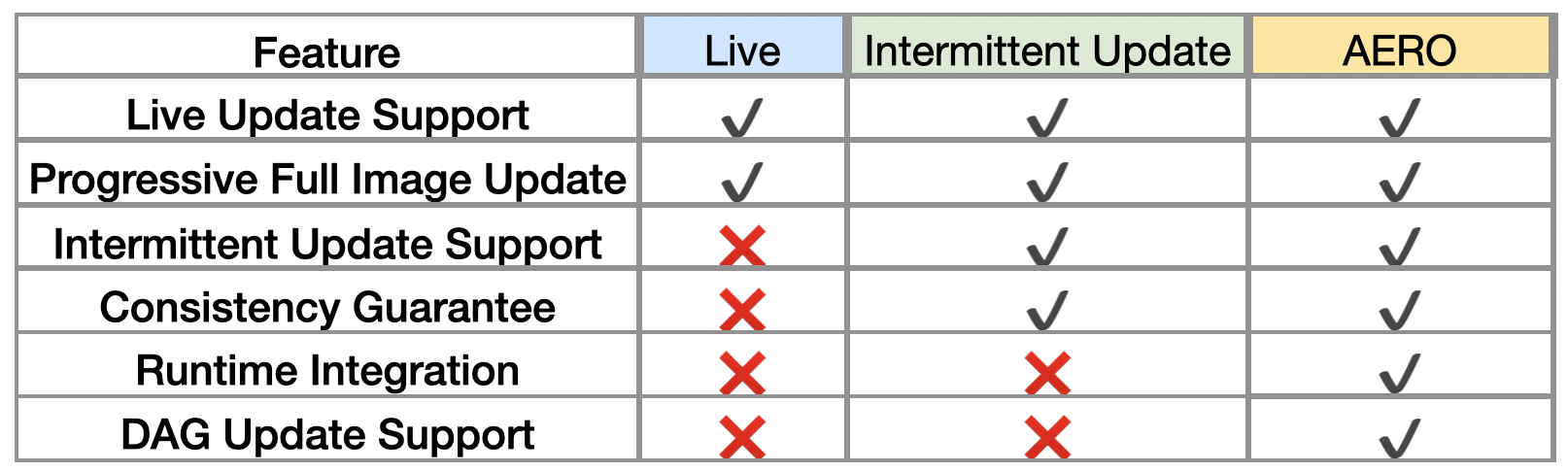}
    \vspace{-6pt}
    \caption{Feature Comparison of Update Approaches}
    \label{fig_func_comp}
\end{figure}

We first compare the features of AERO against the baselines. The baselines offer established capabilities such as live update without reboot and intermittent execution with consistency mechanisms, and both can perform progressive full image update. However, neither supports runtime integration or DAG update. AERO adds these capabilities by integrating update tasks into the DAG and scheduling them alongside routine tasks, enabling task dependency updates that are not possible under existing approaches. Figure~\ref{fig_func_comp} summarizes these differences.

\subsection{Update Error Rate}

\begin{figure}[h]
    \centering
    \includegraphics[width=8cm]{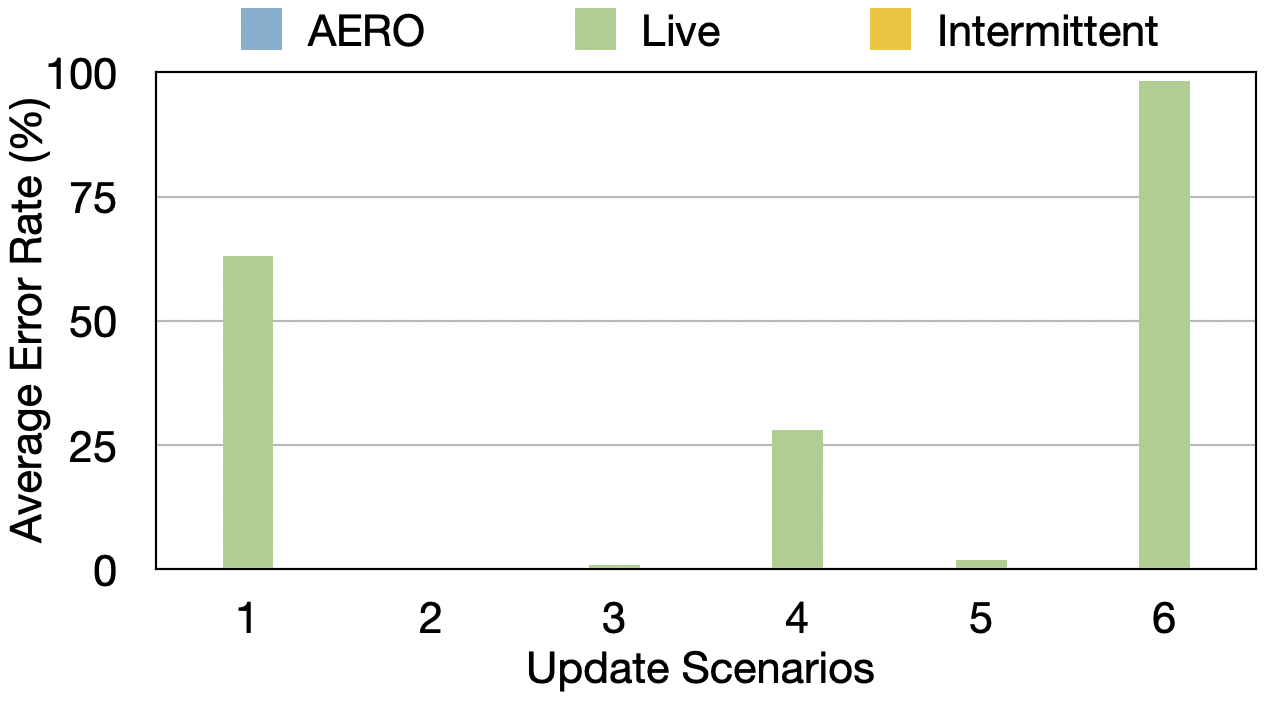}
    \vspace{-6pt}
    \caption{Comparison of Average Error Rate}
    \label{fig_error_rate}
\end{figure}

As shown in Fig.~\ref{fig_error_rate}, AERO maintains a zero error rate across all scenarios. In contrast, the live update approach frequently introduces inconsistencies when updates occur in computation-intensive regions of the DAG; for example, in Scenarios 4 and 6, updates overlap with high-cost tasks, leading to error rates as high as 98.3\% in Scenario 6. Even Scenario 1, which consists of lightweight tasks, shows high error rates under live update because uniformly short execution times increase the chance of interrupting an active task. By comparison, Scenarios 2, 3, and 5 indicate that live update can succeed for small, isolated updates, though additional safeguards are needed to ensure correctness. Intermittent update also maintains a zero error rate by design, but this guarantee comes at the cost of long update delays, which may be unacceptable for time-sensitive updates such as security patches.

\subsection{Update Completion Time}

\begin{figure}[h]
    \centering
    \includegraphics[width=8cm]{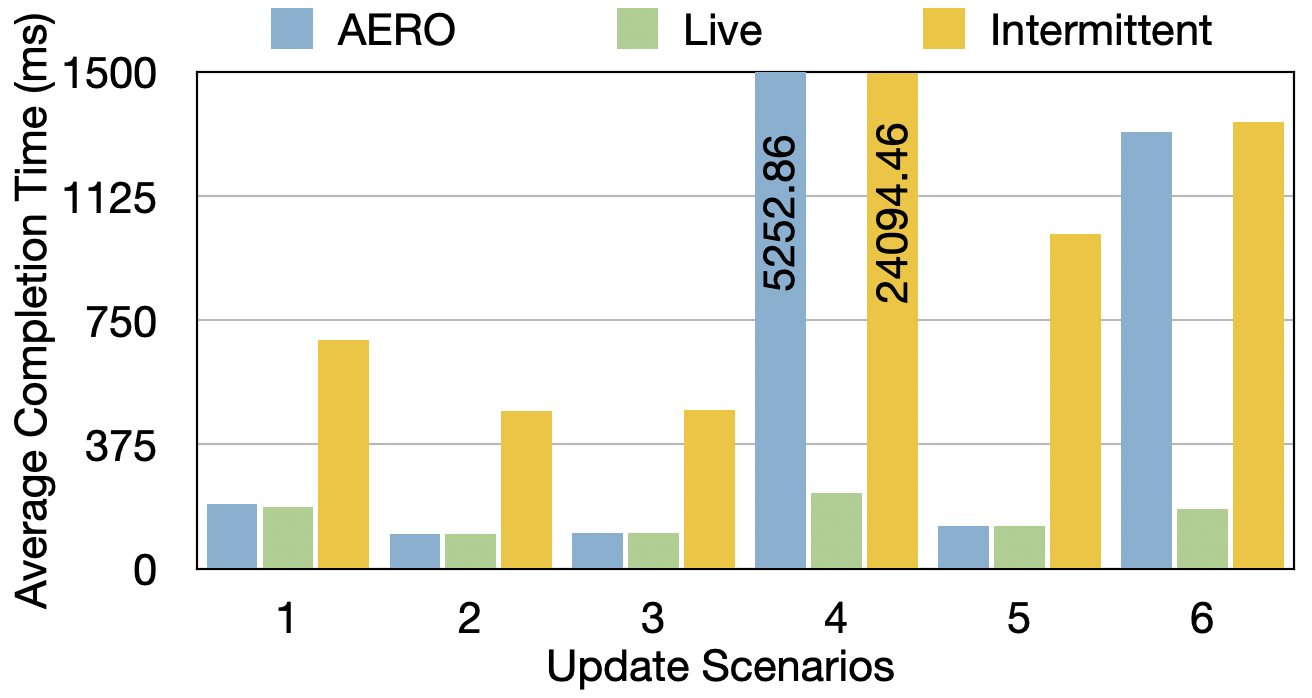}
    \vspace{-6pt}
    \caption{Comparison of Average Update Completion Time}
    \label{fig_update_time}
\end{figure}

Figure~\ref{fig_update_time} shows the average update completion time for all approaches. AERO consistently outperforms intermittent update in Scenarios 1–5 by applying updates more quickly through direct integration of update tasks into the DAG, and even in Scenario 6, which involves the largest update size, it still achieves slightly shorter completion time. Compared to live update, our approach shows similar completion times in Scenarios 1, 2, 3, and 5. Small delays occur only when the \textit{update-affected block} is executing, since updates are deferred to preserve correctness, which aligns with the zero error rate results in Fig.~\ref{fig_error_rate}. The trade-off is most evident in Scenarios 4 and 6, where live update completes faster but suffers from very high error rates, as shown in Fig.~\ref{fig_error_rate}.

\subsection{Deadline Miss Rate (DMR)}

\begin{figure}[h]
    \centering
    \includegraphics[width=8cm]{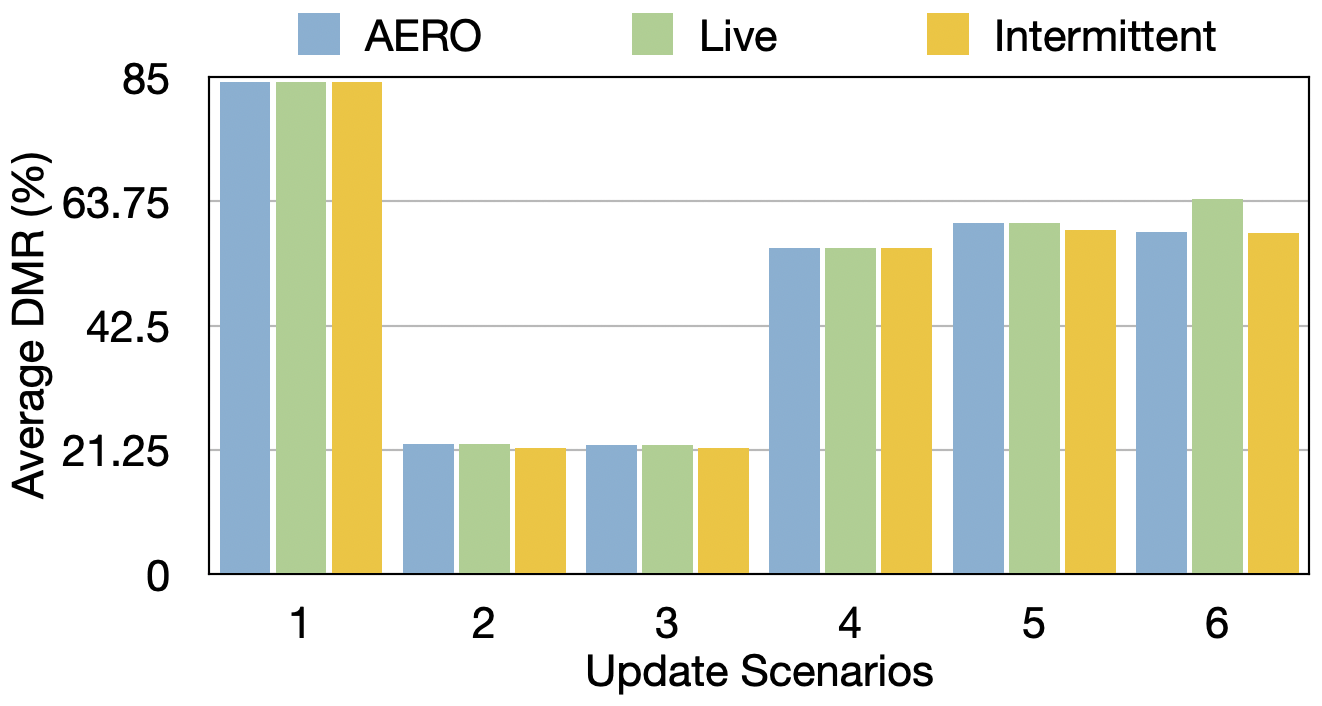}
    \vspace{-6pt}
    \caption{Comparison of Average DMR}
    \label{fig_dmr}
\end{figure}

Figure~\ref{fig_dmr} compares the DMR across the three approaches. Deadlines are defined only for routine tasks as the profiled execution time plus a 0.5× margin to account for scheduling overhead, while update tasks are not assigned deadlines and always execute after routine tasks. AERO achieves results similar to or slightly higher than intermittent update in all scenarios, showing that updates can be integrated at runtime without excessive overhead. Compared to live update, AERO yields similar DMR in Scenarios 1–5. In Scenario 6, where the update size is relatively large, AERO achieves a lower DMR because live update applies updates immediately without considering the current execution state, delaying routine tasks and increasing DMR.

Although our experiments use MSP430FR5994 hardware and solar energy traces, AERO is not limited to this platform or energy source. Its DAG-level design enables broad applicability across diverse EH IoT devices and environments, while introducing overhead only from runtime DAG adjustments triggered upon update arrivals. This overhead primarily incurs non-volatile memory writes for DAG modifications and is negligible in practice for our evaluated scenarios, without affecting the observed evaluation trends.
\section{Conclusion}
\label{conclusion}

This paper presented AERO, a runtime-aware OTA update mechanism for intermittently powered IoT devices. By integrating update tasks into the executing DAG, AERO preserves correctness while enabling updates during normal operation. Simulations using real-world energy traces and hardware profiles show that AERO achieves zero update errors, shorter update completion times, and competitive DMR compared with live and intermittent baselines. Beyond single-device evaluations, AERO also enables future work on coordinated updates in EH IoT networks. Overall, AERO delivers a practical and reliable OTA update solution, with a DAG-level design that generalizes across hardware platforms and EH conditions to support sustainable, long-lived IoT deployments.

\section{Acknowledgment}
This work was supported in part by the U.S. National Science Foundation under Grant CNS-2443885 and CNS-2318641.

\bibliographystyle{unsrt}
\bibliography{ref}

\end{document}